\documentclass[global,final]{svjour}
\usepackage{amsmath,amsfonts}
\usepackage{graphicx}

\def\real{{\rm I\kern-.2em R}}
\def\complex{\kern.1em{\raise.47ex\hbox{  $\scriptscriptstyle
|$}}\kern-.40em{\rm C}}
\def\Box{\hfill\vbox{\hrule height 0.6pt
        \hbox{\vrule width 0.6pt height 1.8ex \kern 1.8ex
                \vrule width 0.6pt}
        \hrule height 0.6pt}}
\title{Adiabatic Theorem without a Gap Condition}
\author{Joseph E. Avron \and Alexander Elgart}
\institute{Department of Physics, Technion, 32000 Haifa, Israel\\
\email{avron@physics.technion.ac.il}} \journalname{cmp}
\date{Received: 3 December 1998/ Accepted: 7 December 1998}
\begin{document}
\flushbottom

\maketitle
\begin{abstract}
We prove the adiabatic theorem for quantum evolution without the
traditional gap condition. All that  this adiabatic theorem needs
is  a (piecewise) twice differentiable finite dimensional spectral
projection. The result implies that the adiabatic theorem  holds
for the ground state of atoms in quantized radiation field.  The
general result we prove gives no information on the rate at which
the adiabatic limit  is approached. With additional spectral
information one can also estimate this rate.
\end{abstract}
\section{Introduction and Motivation}
The adiabatic theorem of Quantum Mechanics  describes the long time behavior of
 solutions of an initial value problem where the Hamiltonian generating the
evolution depends slowly on time.  The theorem relates these
solutions to spectral information of the instantaneous
Hamiltonian.

Traditionally, the adiabatic  theorem is stated for Hamiltonians
that have an eigenvalue which is separated by a gap from the rest
of the spectrum. Folk wisdom is that some form of a gap condition
is a {\it sine qua non} for an adiabatic theorem to hold. This
is based on the following simple but at the same time rather
forceful argument:   The notion of Hamiltonian that depend slowly
on time makes sense provided the system in question has a finite
intrinsic time scale which determines what slow and fast mean.
  In quantum mechanics
the intrinsic time scale  is often
determined by the  gaps in the spectrum (and Planck's constant)
\cite{born}. For
example, a Harmonic oscillator with natural frequency $\omega$, has
gaps in the spectrum whose size is $\hbar\omega$. The condition for
adiabaticity is $|\dot\omega|<<\omega^{2}$. In
the $\omega\to 0$ limit the intrinsic time diverges and
 $\dot\omega\neq 0$
is never adiabatic. This suggests that one can not expect a general
adiabatic theorem to hold in the absence of gaps.

It is, of course, conceivable that in the absence of a gap some
other   property  may determine a relevant and intrinsic time
scale.  For example, in the case of linearly crossing eigenvalues,
the difference in slopes of the eigenvalues at the point of
crossing $ \alpha=(\dot E_{1}-\dot E_{2})$, Fig. 1, determines a
time scale, $\sqrt \frac{1}{\alpha}$, that  takes over as the time
scale associated with the gap diverges. An adiabatic theorem that
builds on this fact goes back to Born and Fock \cite{bf}. But, at
the same time, a general adiabatic theorem in the absence of a gap
condition which does not use some other special properties, like a
slope condition, seems  unlikely and on physical grounds, morally
wrong.

\begin{figure}[htb]
\centering
\includegraphics[height=4.cm]{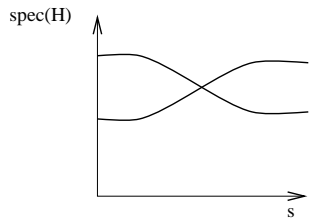}
\caption{Crossing Eigenvalues in Born Fock Theory}
\end{figure}
Nevertheless, the folk wisdom is actually wrong since we shall prove a general
adiabatic theorem without a gap condition. All one really needs
for the adiabatic theorem is a finite dimensional  spectral
projection for the Hamiltonian that depends smoothly on time.
The role of the gap is to
provide an {\em a-priori} {\em rate} at which the adiabatic limit
is approached. In the absence of a gap, there is no such {\em
a-priori} information on the rate at which the adiabatic limit is
approached and it could be arbitrarily slow.

Our approach to an adiabatic theorem without a gap condition has
some of the flavor of an operator analog of the Riemann-Lebesgue
lemma \cite{riemann}. If a function and also its derivative are in
$L^1(\real)$ then it is an elementary exercise that its Fourier
transform  decays at infinity at least as fast as an inverse power
of the argument. The Riemann-Lebesgue lemma says that, in fact,
the Fourier transform of any $L^1(\real)$ function vanishes at
infinity.  The loss of {\it a-priori} information about the
derivative translates to loss of information about the rate at
which the function vanishes at infinity.  In this analogy
differentiability is the analog of the gap condition, and the
$L^1(\real)$ condition is the analog of the smoothness condition
on the spectral projection.

A gap condition is associated with spectral stability.  Situations
without a gap condition often lead to spectral instabilities. This
may suggest that an adiabatic theorem without a gap condition may
be an academic exercise in the sense that it may have no
applications and that its premise, the existence of a smooth
spectral projection, is  either contrived or would be hard to
establish in applications. For example, for applications to atomic
physics, where the essential spectrum is absolutely continuous,
\cite{cycon},  embedded eigenvalues  tend to  dissolve to resonances
\cite{simon} so it is unlikely that the projection
associated to an embedded eigenvalue
 would be continuous.
Indeed, we do not know of
an  application to {\em embedded}
 eigenvalues.

An interesting application of the adiabatic theorem without  a gap condition
is to {\em eigenvalues at threshold}.
A ground state at threshold is a
feature of any reasonable model Hamiltonian  for atoms
interacting with a
radiation field. Models that do not have this property  describe
 unstable atoms, or stable atoms in a world that has no soft
photons.
Models of atom-photon systems have the property that when the fine
structure constant, $\alpha$, is small, the ground state describes the
bound electrons of the atom and a photon field close to the
vacuum. Soft photons are responsible for the absence of a gap in
these models.  A
relatively simple yet interesting model for which the existence
(and uniqueness) of the ground state \cite{s,ah} as well as
gaplessness  \cite{hs} are known rigorously is   the spin-boson
Hamiltonian: The model of a two level system coupled to a
radiation field. This  has also been established
for a model of non-relativistic QED \cite{ah,bfs,bfs1,bfs2,bfss}:
A model of nonrelativistic electrons coupled to a
radiation field with an ultraviolet cutoff. Unfortunately, for
{\em real} QED \cite{bs,cohen}, where both the electrons and photons
are treated as relativistic quantum fields, all that is rigorously
known at present is on a perturbative level.
\begin{figure}[htb]
  \centering
  \includegraphics[height=4.cm]{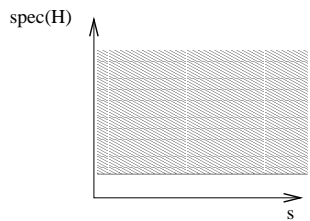}
  \caption{An Eigenvalue at Threshold}
\end{figure}
Our original motivation was  to prove an adiabatic theorem  for
models describing   atom-photon interaction.  We proved  this for
the Dicke model, which is the simplest model of this kind, in
\cite{ae}. We then realized that one could prove a much more
general adiabatic theorem without a gap condition which is not
specific to models of atoms in radiation field, but would cover
these as a special case.

The adiabatic theorem without a gap condition resolves a
 problem regarding the relation between the quantum
mechanics with and without radiation field. If the folk wisdom was
true, and a gap condition was a necessary ingredient in the
adiabatic theorem, one would expect the adiabatic theorem to hold
for a two level system, but not for the spin-boson model. Since
the spin-boson model is clearly a more accurate description of
nature than the model of a two level system, the success of the
adiabatic theorem in numerous applications where a two level model
has been used, would appear like a mystery. The fact that
adiabatic theorems do not really need a gap condition  means that
at least as far as the adiabatic theorem is concerned quantum
mechanics without radiation and quantum mechanics with radiation
sit in the same basket. An interesting problem  that  we
do not resolve here is to show that not only is the adiabatic
theory of quantum mechanics (without radiation) qualitatively
correct, but it is also quantitatively accurate. For the Dicke
model, some results in this direction are given in
\cite{ae}.

\section{Formulation of the Problem and the Main Result}

To formulate the problem of the adiabatic theory
more precisely it is convenient, and traditional, to
replace the physical time $t$ by the scaled time $s=t/\tau$.
One is then concerned with the solution of the initial value problem
\begin{equation} i\, \dot \psi_\tau (s) = \tau H(s)\, \psi_\tau
(s),\label{schrod}
\end{equation}
in the limit of large $\tau$.  $H(s)$ is a
self-adjoint  Hamiltonian which depends sufficiently smoothly
on $s$.
$\psi_{\tau}$ is a vector (in Hilbert space) valued function. We shall be
more specific about what we mean by smoothness below.
 $H(s)$
evolves slowly in  {\em physical} time for a {\em long} interval of time
with {\em finite} variation in $H(s)$.
Quantum adiabatic theorems say that the solution of  the initial
value problem is characterized, in the adiabatic limit $\tau
\to\infty$, by spectral information. There is no single adiabatic
theorem. Different adiabatic theorems focus on different aspects
of the problem:  What is assumed about $H(s)$ and $\dot H(s)$;
about properties of  the projection $P(s)$; the notion of
smoothness, and what are the optimal error estimates, etc. All
have the following structure: Let $P(s)$ be an appropriate family
of spectral projections for $H(s)$.  Let the initial data be such
that $\psi_{\tau}(0)\in Range P(0)$. Then,  for an appropriate
value of $\gamma\ge 0$,
\begin{equation}
dist \Big(\psi_\tau(s), Range P(s)\Big) \le
O\left(\tau^{-\gamma}\right).\label{general}
\end{equation}
For $\gamma=0$ we take the right-hand side to mean
$O\left(\tau^{-0}\right)=o(1)$.

In the present work we shall restrict ourselves to the case where
$\dot H(s)$ is compactly supported. Then we can, without loss,
take $s\in[0,1]$. Second, we shall restrict ourselves to uniform
error estimates, i.e. error estimates that hold for {\em  all}
scaled time. This is actually the easier case. In adiabatic theory
it is often possible to obtain much sharper results for times outside the support of
$\dot H(s)$.

Our main result is the following:
\begin{theorem}\label{main} Suppose that $P(s)$ is smooth finite rank spectral
projection, for the bounded, smooth Hamiltonian $H(s)$. Then, the
evolution of the initial state $\psi_{\tau}(0)\in Range P(0)$,  is
such that $dist\Big(\psi_\tau (s), Range P(s)\Big) \le o(1)$ for
all $s\in [0,1]$.
\end{theorem}
\begin{remark}
This is the weakest, but at the same time, the simplest, and most
characteristic of our results. As it stands, it does not even
apply to the Schr\"odinger operator because $H(s)$ is assumed to
be bounded. In Sect. \ref{unbounded1} we shall state a
generalization of this result to unbounded operators.  There are
 two reasons why we have chosen to state the weaker
result. The first is that we did not want to obscure the central
issue, and  what is new in this work, behind a mask of
technicalities. The second is almost ideological. The adiabatic
problem is an infrared, low energy, problem. The central issue in
an adiabatic theorem without a gap condition is to control low
energy excitations. The unboundedness of Schr\"odinger operators
is an ultraviolet problem. This  problem has well developed
analytical tools
 \cite{kato1,rs,y}, and has nothing to do with the core of the infrared
problem of adiabatic evolution. Once one has an adiabatic theorem
without a gap condition for bounded operators, the extension to
unbounded ones is technical.
\end{remark}
\begin{remark}
We have stated the theorem with a condition of smoothness. Much
less than smoothness is needed and we shall formulate a stronger
result requiring only piecewise, twice differentiability of $P(s)$
in Sect. \ref{unbounded1}. One reason why we have chosen to state
a weaker result is again for simplicity, and the second is that it
is likely that even the  result in Sect. \ref{unbounded1} is not
optimal.
\end{remark}
\begin{remark}
The theorem, as stated, does not cover the case of eigenvalue
crossings. This is because at
 eigenvalue crossing  the spectral projection $P(s)$ is not  smooth ($Tr\,
P(s)$ is discontinuous). Eigenvalue crossings can be handled by a
method due to Kato \cite{kato2} and we shall state a stronger
version of the theorem that allows for finitely many crossings in
Sect. \ref{unbounded1}.
\end{remark}
\subsection{The Results of Davies and Spohn}
Davies and Spohn \cite{ds} studied the evolution of a
driven, finite dimensional quantum system coupled to a heat bath.
Their prime interest was the linear response of
such a system which is closely related to   the adiabatic
limit. They choose a Hamiltonian of the form $$\tau\left(
H_{q}(s) +  H_{f}+ \sqrt\frac{1}{\tau}H_{i}\right),$$ where
$H_{q}(s)$ is the time dependent Hamiltonian of the driven, finite
dimensional, quantum sub-system, $H_{f}$ is the Hamiltonian of a
quasi-free fermion field, and $H_{i}$ is the interaction. The
coupling
vanishes in the adiabatic limit $\tau\to\infty$.
They show that the induced evolution of the finite dimensional
sub-system is governed by a (finite dimensional) Hamiltonian of the form
 $$\tau H_{q}(s) + L(s).$$
Davies and Spohn then proceed to analyze the evolution of this finite
dimensional system using some of
the ideas that enter into the adiabatic theory of Kato \cite{kato2}.
Davies and Spohn do not prove an adiabatic theorem in the sense
that the physical evolution
adheres to a spectral subspace of the coupled Hamiltonian.

\section{A Panorama of Adiabatic Theorems}

In this section we recall some of the basic adiabatic theorems:
Adiabatic theorems with a gap condition, for crossing eigenvalues,
adiabatic theorems beyond all orders, and adiabatic theorems for
scattering. We examine how   these  relate to the
adiabatic theorem without a gap condition.

\subsection{Adiabatic Theorems with a Gap condition.}

The first satisfactory formulation and rigorous proof of an
adiabatic theorem in the then new quantum mechanics was given in
1928 by Born and Fock \cite{bf}. They were motivated by a point of
view advocated by Ehrenfest \cite{ehrenfest}, which identified
classical adiabatic invariants as the observables that get
quantized.  The theorem they proved
was geared to show that quantum numbers are preserved by adiabatic
deformations.

Born and Fock proved an adiabatic theorem for Hamiltonian
operators, $H(s)$, with simple discrete spectrum. They showed that
in Eq.~(\ref{general}) one can take $\gamma\ge 1$. Their proof
covers Hamiltonians like the one dimensional Harmonic oscillator,
but not the Hydrogen atom, which has absolutely continuous
spectrum at positive energies, and eigenvalues with multiplicities
at negative energies.

In 1958 Kato \cite{kato2} initiated a new strategy for proving
adiabatic theorems. He introduced a notion of adiabatic evolution
which is purely geometric. It is associated with a natural
connection in the bundle of spectral subspaces.  Kato's method was
to compare the geometric evolution with the evolution generated by
$H(s)$ and to show that in the adiabatic limit the two coincide.
Using this idea, Kato was able to relax the condition that $H(s)$
had simple discrete spectrum. He showed that the adiabatic theorem
holds when $P(s)$ is  a {\em finite dimensional} spectral
projection associated with an {\em isolated eigenvalue}. No
assumption on the spectral type of $H(s)$ restricted to  $Range
P_{\perp}(s)$ need be made, Fig. 3.

\begin{figure}[htb]
  \centering
  \includegraphics[height=4.cm]{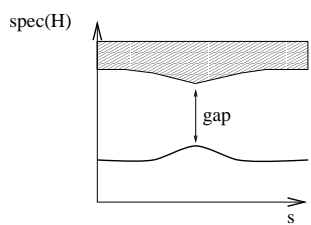}
  \caption{Spectrum in Kato's Theory}
\end{figure}

 Kato's results cover the case of Schr\"odinger operator
 for the Hydrogen atom.  However it does not cover
 Schr\"odinger operators that arise for instance in the study of condensed
matter
 physics, where there is no discrete spectrum at all. Kato's results
 were  extended in \cite{asy,n}, to
 $P(s)$ that need not be associated with an
eigenvalue, and whose rank could also be infinite.
In particular, the initial data could lie
in a subspace corresponding to an energy band provided it
 is separated by a gap from the rest of the spectrum.

\subsection{Adiabatic Theorems beyond all orders.}
There  are interesting and more delicate adiabatic theorems that
apply provided one considers Eq.~(\ref{general}) for times $s$
that lie outside the support of $\dot H(s)$. Assuming a gap
condition and  smoothness (or analyticity)  of $H(s)$
 it has been shown
\cite{g,gp,berry,ks,N} that  the adiabatic theorem
Eq.~(\ref{general}) holds with $\gamma=\infty$. Stronger results
hold in the analytic case \cite{martinez,joye,js,berry}.

\subsection{Adiabatic Theorems with Eigenvalue Crossings}
 Born and
Fock also studied  the adiabatic theorem for crossing  eigenvalues
 where the spectral
projections have smooth continuations  through the crossing point
\cite{bf}.
 Born and Fock
showed that if crossing is of order $m$ (linear crossing is $m=1$)
then  Eq.~(\ref{general})  holds with $\gamma=1/(m+1)$. This
problem was later studied in much detail in
\cite{friedrichs2} and \cite{hagedorn}.

Kato \cite{kato2}  also considered the adiabatic theorems for
crossing eigenvalues. He did not make any explicit assumptions
about how the eigenvalues behave near crossings. The only
assumption he did make was that $P(s)$ could be continued through
the crossings, and that there are finitely
many crossings. Under these conditions he showed that
Eq.~(\ref{general})  holds with $\gamma=0$.

\subsection{Adiabatic Theorems without a Gap Condition}
We are aware of one example of an adiabatic theorem without a gap
condition for operators that have  essential spectrum. This
is a result of \cite{ahs} for rank one perturbations of dense
point spectra. The rate of approach to the adiabatic limit is
$\gamma=1$ in Eq.~(\ref{general}).
\subsection{Adiabatic Theorems for the Scattering Matrix}
Adiabatic scattering theory  relates the time dependent scattering
matrix to the time independent scattering matrix.  Results in this
direction are described in \cite{nt,martinez}. These have very
little to do with the kind of adiabatic theorems we consider here.
In scattering theory a time scale is determined by the initial
data: The scattered particle spends  a finite amount of time in
the region of interaction, and in the limit that the interaction
varies slowly, it does not see the variation in the Hamiltonian.
The adiabatic theorems we are interested in consider a particle
that spends a long time in the region of interaction.

\section{The Adiabatic Theorem and a Commutator Equation}
 In this section we shall describe the proof of Theorem
\ref{main}. To simplify the presentation, we shall stay away from
making optimal assertions. In Sect. \ref{unbounded1} we shall
strengthen the result dropping most of the simplifying
assumptions.

The center of this section, and the heart of the adiabatic
theorem, is the commutator equation, Eq.~(\ref{commutators}). It
is an operator valued equation for two {\em bounded} operators $X$
and $Y$. If one sets $Y=0$ one gets a commutator equation that
goes back to Kato. The commutator equation with $Y=0$ has a
bounded solution $X$ provided there is gap. If there is no gap the
equation may, in some cases, have a bounded solution, but in
general it will not. The basic idea behind the adiabatic theorem
without a gap condition is that one can always solve this equation
with $X$ bounded and $Y$ bounded and small. The smaller $Y$ the
larger is the norm of $X$ in general, but this is all right, as we
shall see.

In this section $H(s)$ is a family of bounded
 self-adjoint Hamiltonians that depends smoothly on $s$ so that $\dot
 H(s)$ is supported in the interval $[0,1]$. $H(s)$
generates unitary evolution as the solution of the initial value problem:
\begin{equation} i\,\dot U_\tau (s) = \tau H(s) U_\tau(s),\quad U_\tau(0)=1,
\quad s\in [0,1].\label{schrodinger}
\end{equation}  We assume, without loss, that $H(s)$ has eigenvalue $0$ and this
eigenvalue has finite multiplicity.  For this eigenvalue we formulate and prove
our main result.

 We recall the notion of  adiabatic evolution
\cite{kato2,asy}. Let $U_A(s)$ be the solution of the initial
value problem:
\begin{equation} i\,\dot U_A(s) =\tau\,\left( H(s)+\frac{i}{\tau}\, [\dot
P(s),P(s)]\right)\, U_A(s),\quad U_A(0)=1, \quad s\in
[0,1].\label{kato2}
\end{equation} It is known that  this  unitary evolution has the
intertwining property \cite{asy}:
\begin{equation}  U_A(s)\, P(0) =P(s)\, U_A(s).
\end{equation} That is, $U_A(s)$ maps $Range\ P(0)$ onto $Range\ P(s)$.
In particular, the solution of the initial value problem
\begin{equation}
i\,\dot \psi(s) = \tau\,\left( H(s)+\frac{i}{\tau}\, [\dot
P(s),P(s)]\right)\,  \psi(s), \ \psi(0)\in Range P(0),
\end{equation}
has the property that $\psi(s)\in Range P(s)$. We shall show that the
Hamiltonian evolution,
$U_\tau(s)$, is close to the adiabatic evolution  $U_{A}(s)$.

 We first formulate the basic lemma:
\begin{lemma} Let  $P(s),\ s\in[0,1],$ be a
differentiable  family of spectral projections for the self-adjoint Hamiltonian
$H(s)$ with (operator) norm
$\Vert
\dot P(s)\Vert<\infty$. Suppose that the commutator equation
\begin{equation} [\dot P(s), P(s)] = [H(s),X (s)]
+Y(s)\label{commutators}
\end{equation} has operator valued solutions, $X (s)$ and $Y (s)$ with
$X (s)$, $\dot X (s)$ and $Y (s)$ bounded.
Then
\begin{eqnarray}
& \Vert (U_\tau(s)-U_A(s))\,P(0)\Vert \le \nonumber\\ & \max_{s\in
[0,1]} \left( \frac{2\,\Vert X (s)\,P(s)\Vert+\left\Vert \dot{
\Big(X(s) \,P(s)\Big)}P(s)\right\Vert }{\tau}+\Vert \,Y (s)
\,P(s)\Vert\right).\label{estimate}
\end{eqnarray}

\end{lemma}
 The commutator equation, Eq.~(\ref{commutators}), can be
viewed as a definition of $Y(s)$. The issue is not to find a solution to
this equation, but rather to find solutions that make $Y$
small.
In the case that there is a gap $\Delta$ separating
the eigenvalue from the rest of the spectrum, a solution of the
commutator equation is
\begin{equation}
X(s)=\frac{1}{2\pi i}\,\int_\Gamma  \,R(z,s)\, \dot P(s) R(z,s) \,
dz,\quad Y(s)=0.\label{gap}
\end{equation}
Here $\Gamma$ is a circle in the complex plane, centered at the eigenvalue,
 and of
radius $\Delta/2$, Fig. 4. $R(z,s)=(H(s)-z)^{-1}$ is the resolvent
at scaled time $s$.  In this case the  rate  at which the
adiabatic limit is obtained, is seen from  Eq.~(\ref{estimate}) to
be  $1/\tau$.

\begin{figure}[htb]
  \centering
  \includegraphics[height=4.5cm]{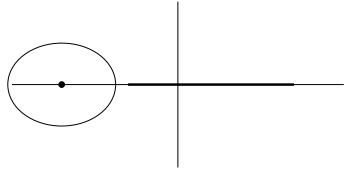}
  \caption{A contour $\Gamma$ in the Complex Plane}
\end{figure}
The strategy for proving the adiabatic theorem without a gap
condition is to show that one can pick $Y$ so that its norm is
arbitrarily small, possibly at the expense of large norm for $X$
and $\dot X$. So long as the norm of $X$ and $\dot X$ is finite,
it can be compensated by taking $\tau$ large. This means that one
can make the right-hand side of Eq.~(\ref{estimate}) arbitrarily
small.  The price paid is that there is, generally speaking, no
information about the rate at which the adiabatic limit is
obtained.

\begin{proof}
 Let $W(s) =  U_\tau^\dagger(s) U_A(s)$ be the wave operator comparing the
adiabatic and Hamiltonian evolution. Since
\begin{equation}
\Vert U_\tau(s)-U_A(s)\Vert=\left\Vert
U_\tau(s)\Big(1-W(s)\Big)\right\Vert=\Vert 1-W(s)\Vert,
\end{equation} we need to bound $W(s)-1.$ From the definition of the
adiabatic evolution, the commutator equation, and the equation of
motion
\begin{eqnarray}
\dot W(s)&=&U^\dagger_\tau(s)\left( [\dot P(s),P(s)]
\right)\,U_A(s)\nonumber \\ &=& U^\dagger_\tau(s)\left( [\dot
P(s),P(s)] \right)\,U_\tau(s)\,W(s) \nonumber \\ &=&
U^\dagger_\tau(s)\Big( [ H(s),X(s)] +Y(s) \Big)\,U_\tau(s)\,W(s)
\\
&=& -\frac{i}{\tau}\Big(\dot U^\dagger_\tau(s)X(s)U_\tau(s) +
U^\dagger_\tau(s)X(s)\dot U_\tau(s)\Big)\,W(s)\nonumber \\ &+&
U^\dagger_\tau(s) Y(s)U_A(s)\nonumber \\ &=&
-\frac{i}{\tau}\Big(\dot{( U^\dagger_\tau(s)X(s)U_\tau(s))} -
U^\dagger_\tau(s)\dot X(s)U_\tau(s)\Big)\,W(s)\nonumber \\ &+&
U^\dagger_\tau(s) Y(s)U_A(s) \nonumber \\ &=&
-\frac{i}{\tau}\left\{\dot{\Big(
U^\dagger_\tau(s)X(s)U_\tau(s)\,W(s)\Big)} - U^\dagger_\tau(s)
X(s)U_\tau(s)\,\dot W(s)\right.\nonumber \\ &&\left
.\phantom{\frac{i}{\tau}}- U^\dagger_\tau(s)\dot
X(s)U_\tau(s)\,W(s)\right\}+\,U^\dagger_\tau Y(s)U_A(s)\nonumber
\\
&=& -\frac{i}{\tau}\left\{\dot{\Big(
U^\dagger_\tau(s)X(s)U_A(s)\Big)} - U^\dagger_\tau(s) X(s)\,[\dot
P(s),P(s)]\,U_A(s)\right .\nonumber
\\&&\left.\phantom{\frac{i}{\tau}}- U^\dagger_\tau(s)\dot
X(s)U_A(s)\right\}+\,U^\dagger_\tau(s) Y(s)U_A(s)\nonumber.
\end{eqnarray}
 The lemma then follows by integration since $W(s)$ is unitary with
$W(0)=1$.\qed
\end{proof}
Let us  describe a solution of the commutator equation which is
motivated by the solution Eq.~(\ref{gap}) in the case of a gap. In
order to have explicit error estimates and also in order to make
the presentation simple and as elementary as possible, we choose a
Gaussian regularizer.
\begin{definition}
Let $g$ and $e$ denote the Gaussian and Error functions\footnote{The
error function we use differs by a  factor and shift from
canonical error function.}, and $\Phi$ be
the special function defined below:
\begin{equation}
g(\omega)= e^{-\pi\omega^{2}},\quad e(\omega)=\int_{-\infty}^{\omega}ds\,
g(s),\quad \Phi(\omega)=\theta(\omega)- e(\omega),
\end{equation}
$\theta$ is the usual step function which vanishes for negative
argument. Also, let us denote the scaling of a function by
\begin{equation}
g_{\Delta}(\omega)=g(\Delta \omega),
\end{equation}
and the multiplication operator by the argument by
\begin{equation}
(\omega g)(\omega)=\omega \,g(\omega).
\end{equation}
\end{definition}
An elementary lemma is:
\begin{lemma}
$\Phi$ has finite $L^{1}$ norm and finite moments. In particular:
\begin{equation}
\Vert \Phi \Vert_{1} =  \frac{1}{\pi},\quad \Vert \omega\Phi \Vert_{1}=
 \frac{1}{4\pi}.
\end{equation}
Under scaling, $\Delta>0$:
\begin{equation}
\left\Vert \Phi_{\Delta} \right\Vert_{1} = \frac{1}{\pi\Delta},
\quad \left\Vert
\omega\Phi_{\Delta}\right\Vert_{1}=\frac{1}{4\pi\Delta^{2}}.
\end{equation}
\end{lemma}
We assume, without loss, that the spectral projection $P(s)$ is
associated with the eigenvalue zero.
\begin{lemma}\label{estimates}
Let $P(s)$ be  a smooth spectral projection for $H(s)$ associated
with the eigenvalue zero. Let $\Gamma$ be an infinitesimal contour
around the origin in the complex plane.\footnote{The choice of
Gaussian is not optimal.  It would be   more convenient to choose
a regularizer which is a better approximant to a characteristic
function and  the reader may want to think of a Gaussian which is
flattened at the top.}
 Then the commutator
equation has the solution
\begin{eqnarray}
X_\Delta(s)&=& A+A^\dagger,\quad A=P(s)\, \dot P(s)
R(0,s)\,\left(1-g\left(\frac{H(s)}{\Delta}\right)\right);\nonumber
\\ Y_\Delta(s)&=& -g\left(\frac{H(s)}{\Delta}\right)\dot P(s)
P(s)+P(s)\dot P(s) g\left(\frac{H(s)}{\Delta}\right),\label{XY}
\end{eqnarray}
where \begin{eqnarray}\label{normY}
 \Vert X_\Delta(s)P(s)\Vert &\le
& \frac{2\Vert \dot P(s)P(s) \Vert}{\Delta},\\
 \Big\Vert \dot {\Big(X_\Delta(s)P(s)\Big)}\Big \Vert&\le& \frac{2(\Vert
\ddot P(s)
 \Vert +\Vert
\dot P^{2}(s))
 \Vert}{\Delta}+
  \frac{\pi\,\Vert \dot P(s)
\Vert\, \Vert \dot H\Vert }{\Delta^{2}}.
\end{eqnarray}
\end{lemma}
\begin{proof}
We start with a formal calculation. Let
\begin{eqnarray}
&F_\Delta(s)=g\left(\frac{H(s)}{\Delta}\right)-P(s)\,,\nonumber\\&
X_\Delta(s)=\frac{1}{2\pi i}\,\int_\Gamma  dz\,
(1-F_\Delta(s))\,R(z,s)\, \dot P(s) R(z,s)\,(1-F_\Delta(s)).
\end{eqnarray}
Since $\dot P(s) = P(s)\dot P (s) +\dot P (s)P(s)$,
$X_\Delta (s)$ can be written as a sum of two adjoint terms, one of them is
\begin{eqnarray}
\frac{1}{2\pi i}\,&\int_\Gamma& dz\, (1-F_\Delta(s))\,R(z,s)\,P(s) \dot P(s)
R(z,s)\,(1-F_\Delta(s))\nonumber\\&=&
\frac{1}{2\pi i}\, (1-F_\Delta(s))\,P \dot P(s)
\left(\int_\Gamma dz\,\frac{R(z,s)}{z}\right)\,(1-F_\Delta(s))\nonumber \\&=&
\,P(s) \dot P(s)
{R(0,s)}(1-P(s))\,(1-F_\Delta(s))\nonumber \\&=&
  P(s) \dot P(s)
{R(0,s)}(1-P(s)-F_\Delta(s))\nonumber \\&=&
 P(s) \dot P(s)
{R(0,s)}\left(1-g\left(\frac{H(s)}{\Delta}\right)\right)
=A.
\end{eqnarray}
We have used
\begin{equation}
P(s)\,F_\Delta(s)=F_\Delta(s)\,P(s)=\Big(g(0)-1 \Big)\,P(s)=0.
\end{equation}
Using this integral representation of $X_\Delta (s)$ we now find $Y_\Delta
(s)$. By our choice of
$F_\Delta(s)$ we have $[F_\Delta(s),H(s)]=0$.
Hence,
\begin{eqnarray}
 &&[X_\Delta(s),H(s)]\nonumber \\
 &=& \frac{1}{2\pi i}\,\int_\Gamma\,
\Big[ (1-F_\Delta(s) )\, R(z,s)\, \dot P(s) R(z,s)
\,(1-F_\Delta(s)), H(s)-z\Big]\,dz \nonumber
\\& =&\frac{1}{2\pi i}\,\int_\Gamma\,dz\,(1-F_\Delta(s))\Big[
\,R(z,s)\, , \dot P(s)\Big](1-F_\Delta(s)) \nonumber \\ &
=&(1-F_\Delta(s))\,\Big[ P(s) ,\dot P(s) \Big]\,
(1-F_\Delta(s))\nonumber \\ & =&[ P(s) ,\dot P(s)]
-\Big\{F_\Delta(s),[ P(s), \dot P(s)]\Big\} +
F_\Delta(s)[P(s),\dot P(s)]F_\Delta(s)\nonumber
 \\
 & =&[ P(s)
,\dot P(s)] + F_\Delta(s)\dot P(s) P(s)-P(s)\dot P(s) F_\Delta(s).
\end{eqnarray}
So a solution of the commutator equation is
\begin{equation}
Y_\Delta(s)= -g\left(\frac{H(s)}{\Delta}\right)\dot P(s)
P(s)+P(s)\dot P(s) g\left(\frac{H(s)}{\Delta}\right).\label{Y}
\end{equation}
It remains to estimate the norms of $X$ and $\dot X$. Using the
fact the Gaussian is its own Fourier transform,\begin{equation}
g\left(\frac{H(s)}{\Delta}\right)=\Delta \int_\real g(\Delta\,t)\,
\exp [{2\pi i t H(s)}] dt,
\end{equation}
one  checks that with our choice of $\Phi$
\begin{equation}
 R(0, s) \left(1-g\left(\frac{H(s)}{\Delta}\right)\right)=
 2\pi i\,\int_\real \Phi\left({t}{\Delta}\right)\, \exp
[{2\pi i t
H(s)}] \,dt.
\end{equation}
Hence
\begin{equation}
\left\Vert R(0, s) \left(1-g\left(\frac{H(s)}{\Delta}\right)\right)
\right\Vert\le
2\pi\,\left\Vert \Phi_{\Delta}\,\right\Vert_1=
 \frac{2}{\Delta}.
\end{equation}
Using the equation for $X(s)$ this estimate proves the bound on
$X(s)$. To get a bound on $\dot {\Big(X_\Delta(s)P(s)\Big)}$,  use
the Duhammel formula,
\begin{equation}
\dot{\Big(\exp (2\pi i t H(s))\Big)}= 2\pi i \, t\, \int_{0}^{1} dz\,e^{
2\pi iz t
H(s)}\, \dot H(s) \,e^{ 2\pi i(1-z) t
H(s)}.
\end{equation}
Collecting the various terms give the claimed estimate. \qed
\end{proof}
 As Lemma \ref{estimates} shows, as $\Delta$ shrinks, the norms of
$X(s)$ and $\dot X(s)$ may, and in general, will, grow. This,
however is of no concern, as long as the norms remain finite, for
one can always compensate for this growth by choosing $\tau$ large
enough. The good thing about shrinking $\Delta$ is that this can
be used to make the norm of $Y_\Delta$ small. Hence, we can always
make the right-hand side of Eq.~(\ref{estimate}) arbitrarily
small.
\begin{lemma}
Suppose that   $H(s)$ is smooth with a zero eigenvalue with
spectral projection $P(s)$ smooth and of finite rank. Let
$g\left(\frac{H(s)}{\Delta}\right)$ be as above. Then $\Vert
Y_\Delta(s)\,P(s)\Vert=\Vert g\left(\frac{H(s)}{\Delta}\right)\dot
P(s)P(s)\Vert \to 0$ uniformly as  $\Delta$ shrinks to zero.
\end{lemma}
\begin{remark}
We owe the proof below to Michael  Aizenman.
\end{remark}
\begin{proof}
For the sake of simplicity suppose that $P(s)$ is a
one-dimensional projection with $P(s)\psi(s)=\psi(s)$,  $\psi$  is
normalized to 1. Let $\varphi=\dot P(s)\psi $. Then, using
$P(s)\varphi(s)=P(s)\dot P(s)\psi(s)=P(s)\dot P(s)P(s)\psi(s)=0$,
we obtain
\begin{eqnarray}
\Vert g\left(\frac{H(s)}{\Delta}\right)\dot P(s) P(s)\Vert^2 =
\Vert g\left(\frac{H(s)}{\Delta}\right) \dot P(s) \psi(s)\Vert^2 =
\Vert g\left(\frac{H(s)}{\Delta}\right)
\varphi(s)\Vert^2&=&\nonumber \\ \left\langle  \,\varphi
\,|g^2\left(\frac{H(s)}{\Delta}\right)
 | \,\varphi \, \right\rangle=
 \int_{\sigma(H(s))}g^2(x/\Delta)d\mu_\varphi(x),
\end{eqnarray}
where $\mu_\varphi$ denotes the spectral measure.
Now, $g\left(\frac{x}{\Delta}\right) $ is bounded by one, and  goes
monotonically
to zero for all $x\neq 0$, and $g(0)=1$. Hence
\begin{equation}
\lim_{\Delta\to 0}\int_{\sigma(H(s))}g^2(x/\Delta)d\mu_\varphi(x)=
\mu_\varphi(0)=0.\label{measure}
\end{equation}
 It follows that
there is a sequence of $\Delta$ that makes $Y_\Delta (s)$
arbitrarily small.\qed
\end{proof}
This completes the proof of Theorem \ref{main}.

The physical interpretation of the  adiabatic theorem without a gap
condition is that although
the adiabatic theorem ``always'' holds, it does so for different physical
mechanisms. In the case that there is a gap in the spectrum the
adiabatic theorem holds because the eigenstate is protected by a gap from
tunneling out of the spectral subspace. In the case that there is no gap
and the
spectrum near the relevant eigenvalue is essential, the adiabatic theorem holds
because essential spectrum is associated with states
that are supported near spatial infinity. There is little tunneling
to these states because of small overlap with the wave function
corresponding to an eigenvalue which is  supported away from infinity.

\section{Fine Print}\label{unbounded1}
In this section we extend the adiabatic theorem without a gap
condition to unbounded self-adjoint operators; replace the
smoothness condition  by a condition on differentiability and allow
 eigenvalue crossing.
These extensions are technical in character and rely on existing
machinery.
\subsection{Unbounded Hamiltonians}
The first, and perhaps the main, difficulty with unbounded
operators $H(s)$ is the existence of solutions to  the initial
value problem, Eq.~(\ref{schrod}). For bounded operators the
existence is a  consequence of the Dyson formula, see e.g. Theorem
X.59 in \cite{rs}. For unbounded operators existence is more
subtle so we chose a class for which this is the case:
\begin{definition}
A family of (possibly unbounded)  self-adjoint Hamiltonians $H(s)$
is admissible if
\begin{enumerate}
\item $H(s)$ have the common domain in Hilbert space for all
$s\in[0,1]$.\label{domain}
\item $H(s)$ is bounded from below by $\Lambda$.
\item $R(i,s)$ is bounded and differentiable  and
$H(s)\dot R(i,s)$ is bounded.\label{R}
\end{enumerate}
\end{definition}
It is a consequence of our definition of admissibility that
$A(t)=(H(t)-\Lambda+1)$  is a strictly positive operator.
Moreover, it is follows from property (\ref{domain}) by a closed
graph theorem that $A(t)A(s)^{-1}$ is bounded. Since, for $t-s$
small, $\Vert(t-s)^{-1}(A(t)A(s)^{-1}-I)\Vert=\Vert
A(s)\dot{A}(s)^{-1}\Vert+o(|t-s|)$, the last expression is bounded
due to property (\ref{R}). The existence of the unitary evolution
for an admissible family of Hamiltonians follows now from
(\cite{rs}, Theorem X.70):
\begin{theorem}\label{existence}
Let $X$ be a Banach space and let $I$ be an open interval
 in $\mathbb{R}$. For each $t\in I$, let
$A(t)$ be the generator of a contraction semigroup on
$X$ so that $0\in \rho(A(t))$ and
\begin{enumerate}
\item The $A(t)$ have the common domain $D$.
\item For each $\phi\in X$, $(t-s)^{-1}(A(t)A(s)^{-1}-I)\phi$ is uniformly
strongly
continuous and uniformly bounded in $s$ and $t$ for $t\neq s$ lying in any
fixed compact
subinterval of $I$.
\item For each  $\phi\in X$, $C(t)\phi\equiv \lim_{t\rightarrow
s}(t-s)^{-1}(A(t)A(s)^{-1}-I)\phi$
 exists uniformly for $t$ in each compact subinterval and $C(t)$ is bounded
and strongly continuous in $t$.
\end{enumerate}
Then unitary evolution exists uniformly in $s$.
\end{theorem}

Then we can prove the following result.
\begin{theorem}\label{unbounded} Suppose that $P(s)$ is finite rank
spectral projection which is at  least twice differentiable (as a
bounded operator), for an admissible family $H(s)$. Then, the
evolution of the initial state $\psi(0)\in Range P(0)$, according
to Eq.~(\ref{schrod}), is such that for all $s\in [0,1]$,
$dist\Big(\psi_\tau (s), Range P(s)\Big) \le o(1)$.
\end{theorem}
\begin{proof}
Tracing the steps in  Theorem~\ref{main} one  sees that it is
enough to check that the operators $X(s)$, $\dot X(s)$ and
 $Y(s)$ are bounded uniformly in $s$. Now, by Eq.~(\ref{XY}),
 $X_{\Delta}$ and $Y_{\Delta}$ are made of  bounded
 operators such as $P$, $\dot P$ and $R$. Moreover,  $X_{\Delta}$ is
 also differentiable as a bounded operator by our assumption that $P$
 is twice differentiable, and by the admissibility condition that
 guarantees that $R$ is differentiable as a bounded operator.  By the
functional calculus  $g(H(s))$ is also differentiable as a
 bounded operator. The only change is  in the explicit
 estimate on the norm of $\dot X_{\Delta}$ in terms of $\Delta$,
 which is replaced by
 \begin{eqnarray}
 \Big\Vert \dot {\Big(X_\Delta(s)P(s)\Big)}\Big \Vert&\le& \frac{2(\Vert
\ddot P(s)
 \Vert +\Vert
\dot P^{2}(s))
 \Vert}{\Delta}\nonumber\\ &+&
  \frac{\pi\,\Vert \dot P(s) (H(s)+i)
\Vert\, \Vert R(s,i)\,\dot H\Vert }{\Delta^{2}},
\end{eqnarray}
which is bounded for admissible $H(s)$.\qed
\end{proof}
\subsection{Piecewise Differentiability and Eigenvalue Crossing }
If at some time $0<s_0<1$  crossing of eigenvalues occurs, then the
spectral projection associated with one of the eigenvalues,
$P(s)$, is discontinuous at $s_0$ since its rank jumps.
Suppose that $P(s), \ s\neq s_{0}$ is a spectral projection whose
 limit from the right and left coincide at
$s_{0}$. In this case we can use an argument of Kato \cite{kato2}
that shows that global continuity together with piecewise
smoothness is good enough.

Kato's argument goes as follows: Choose a small $\varepsilon$. The
physical evolution follows the adiabatic evolution up to an
arbitrarily small error on
 the interval $[0,s_{0}-\varepsilon]$.  On the short
interval  $[s_{0}-\varepsilon,s_{0}+\varepsilon]$ the physical
evolution is $\varepsilon$  takes $Range P(s_{0}-\varepsilon)$
close to itself. Since $P(s)$ is continuous at $s_{0}$, by
assumption, this is equivalent to the statement that the physical
evolution takes $Range P(s_{0}-\varepsilon)$ close to $Range
P(s_{0}+\varepsilon)$, with an error that can be made arbitrarily
small with $\varepsilon$. The physical evolution now follows the
adiabatic evolution up to an arbitrarily small error on
 the interval $[s_{0}+\varepsilon,1]$. Summarizing we have:
\begin{theorem}\label{crossing} Suppose that ${P}(s), \ s\neq
s_{0}\in[0,1]$, is a  finite rank spectral projection which is
piecewise twice differentiable (as a bounded operator) and is
everywhere continuous on $[0,1]$. Then the initial data
$\psi_{\tau}\in Range P(0)$ evolve according to Eq.~(\ref{schrod})
so that $dist\Big(\psi_\tau (s), Range P(s)\Big) \le o(1)$ for all
$s\in [0,1]$.
\end{theorem}
\section{The  Rate of Approach to the Adiabatic Limit}
The general adiabatic theorems we have formulated give no
information on the rate at which the adiabatic limit is
approached.  In fact, from the results of Born and Fock and Kato
about eigenvalue crossings, it is clear that in the absence of a
gap, the rate can be arbitrarily slow. To get interesting results
on  the rate at which the adiabatic limit is approached
necessarily involves additional spectral information. In
particular, if the bound state is either embedded or at the
threshold of essential spectrum, with good behavior
 of the spectral measure at nearby energy, one expects to do better.
  An illustration of such estimates is
given below.

Recall \cite{last} that a (Borel) measure $\mu$ is called
(uniformly) $\alpha$-H\"older continuous, $\alpha\in[0,1]$, if
there is a constant $C$ such that for every interval $\Delta$ with
$|\Delta|<1 $, \footnote{$|\cdot|$ denote
 Lebesgue measure.}
\begin{equation}
\mu(\Delta)<C|\Delta|^\alpha.
\end{equation}
The interest in such measures comes from the fact
\cite{last,combes,guarneri} that they carry dynamical information
and  $\alpha$-continuous measures are the limits of
$\alpha$-H\"older continuous measures. Knowing something about the
Hausdorff dimension of the spectrum \cite{last} then translates to
information about the rate of approach of the adiabatic limit.

\begin{corollary}\label{dos}
If the spectral measure $\mu_\varphi(\Delta)$, is
$\alpha$-H\"older continuous then the adiabatic limit is
approached at least at rate
$\gamma=\frac{\alpha}{2+\alpha}$. In the case
of a family of Hamiltonians related by unitaries,
$H(s)= V(s) HV^{\dagger}(s)$, with $\dot V V^{\dagger}$ bounded and
differentiable
 the rate is at least
$\gamma=\frac{\alpha}{1+\alpha}$.
\end{corollary}
\begin{proof}
Let us note, first of all, that if the spectral measure
$\mu_\varphi(\Delta)$, is $\alpha$-H\"older continuous
 then rhs of~(\ref{measure})
 is bounded by $\tilde C\vert \Delta\vert^\alpha$. Indeed,
\begin{eqnarray}
\int_{\sigma(H(s))}g^2(x/\Delta)d\mu_\varphi(x) &<&
\sum_{n=0}^{\infty}
g^2(n)(\int_{n\Delta}^{(n+1)\Delta}d\mu_\varphi(x)+
\int_{-(n+1)\Delta}^{-n\Delta}d\mu_\varphi(x)) \nonumber \\ &<&
C\vert \Delta\vert^\alpha \sum_{n=-\infty}^{\infty}g^2(n)= \tilde
C\vert \Delta\vert^\alpha.
\end{eqnarray}
Collecting the various error estimates one gets for the right hand
side of Eq.~(\ref{estimate}) the upper bound
\begin{equation}
\frac{A}{\Delta \tau}+\frac{B}{\Delta^2 \tau} + \tilde C\vert
\Delta\vert^\alpha.
\end{equation}
$A,B$ and $C$ are constants. For the case of the family of
Hamiltonians related by unitaries, by Eq.(\ref{normX}) below,
$B=0$. Optimizing the choice of $\Delta$ gives the result. \qed
\end{proof}
 \subsection{Unitary Families}
By unitary families we mean the special case where the family $H(s)$
has the form
 \begin{equation}
H(s)=V(s) H V^{\dagger}(s),\end{equation} with $V(s)$ unitary.
There are three points that we want to make about unitary
families. The first is that such families are interesting in the
context of adiabatic dynamics from the perspective of
applications. The second is that there is some simplification that
occurs for such families.

In a moving frame, the
Schr\"odinger equation, Eq.~(\ref{schrod}), for $\psi_{\tau}=
V\phi_{\tau}$ takes the form:
 \begin{equation}
i\dot \phi_{\tau}= \tau \left(H+ \frac{i}{\tau}V^{\dagger}(s)\dot
V(s)\right)\phi_{\tau}.\end{equation} This
leads to time independent Hamiltonian in the very special case:
$V(s)=e^{is\sigma}$, with $\sigma$
self-adjoint (fixed) operator.
 The general case
of unitary families, even in the rotating frame, leads to  a time
dependent problem, albeit one with a weak time dependent
perturbation. As this perturbation is allowed to act for a long
time, there is no obvious simplification in the rotating frame.

Unitary families often enter in applications. See for example, M.
Berry's model of a spin half in a magnetic field \cite{berry}
$$H(s)= B(s)\cdot\sigma,$$ where $\sigma$ is a vector of Pauli
matrices, and $B(s)$ a vector in $\real^{3}$ of unit length. This
is a unitary family, which
 has all the intricacies of adiabatic theory
 associated with e.g. Zener tunneling \cite{js}.

Now we come to the simplification. In the case of unitary families
one can improve the estimate of the norm  of $\dot X$, which affects
the estimate of the rate $\gamma$.

In the case of a unitary family
\begin{equation}
P(s) =V(s) PV^{\dagger}(s),\quad \dot P(s) =[\dot V(s)
V^{\dagger}(s), P(s)].\label{rotation}
\end{equation}
Hence,  applying Eq.~(\ref{rotation}) to $X_{\Delta}(s)
=A(s)+A^{\dagger}(s)$, where $A,A^\dagger$ are given by
Eq.~(\ref{XY}), we derive
\begin{equation}
A(s)= V(s) P  [ V^{\dagger}(s) \dot V(s), P]
R(0,0)\left(1-g\left(\frac{H}{\Delta}\right)\right)V^{\dagger}(s).
\end{equation}
Therefore,
\begin{eqnarray}
\dot A(s)= [\dot V(s) V^{\dagger}(s),A(s)]+
 V(s) P  [\dot{ V^{\dagger}(s) \dot V(s)}, P]
R(0,0)\cdot\nonumber\\ \cdot
\left(1-g\left(\frac{H}{\Delta}\right)\right)V^{\dagger}(s).\label{normX}
\end{eqnarray}
This identity is the reason  why, for unitary families, one gets an
improved
rate with $\gamma= \frac{\alpha}{\alpha+1}$.
\subsection {Friedrichs Models}
H\"older continuity of the spectral measure gave an estimate of the
 rate $\gamma= \frac{\alpha}{\alpha+2}\le \frac{1}{3}$, in the general
 case and
 $\gamma= \frac{\alpha}{\alpha+1}\le \frac{1}{2}$
 in the case of unitary families. Presumably,
neither is optimal, since we used the
additional spectral information only to estimate the norm of
$Y_{\Delta}$, but  not to improve the estimate on $X_{\Delta}$ and
$\dot X_{\Delta}$. As a
consequence, the best rate we get is $\gamma=\frac{1}{2}$.
It is
intriguing that for classical  ergodic systems the approach to the
adiabatic limit in the {\em classical adiabatic theorem} is with
 rate $\gamma=\frac{1}{2}$  \cite{ott}.  This
does not imply that the rate of approach to the adiabatic
limit must be slow compared to the rate with a gap.
  In this subsection we shall consider a class of models,
patterned after Friedrichs \cite{friedrichs1}, where a more
precise estimate of $\gamma$ can be made and where $\gamma$ can
also take the value $1$ in the absence of a gap.

Let us consider the family of unitarily related Hamiltonians
$H(s)= V(s) H V^{\dagger}(s)$. At any given time, $s$, there
exists a representation of the Hilbert space such that ${\cal
H}_s= \complex\oplus L^2(\real^d,d\mu(k))$ with $inf(support\,
\mu)>-\infty$ and $ \mu (0)=0$. A vector $\Psi\in{\cal H}$ is
normalized by
\begin{equation}
\Psi=\left(\begin{array}{c} \omega\\ \psi(k)
\end{array}\right), \quad \Vert \Psi\Vert^2= |\omega\vert^2
+\int_{\real^d} \vert \psi(k)\vert^2 d\mu(k),\quad
\omega\in\complex.
\end{equation}
The (Friedrichs)  Hamiltonian $H(s)$ in this representation acts
 on ${{\cal{H}}_s}$ like so:
\begin{equation} H(s)\,\Psi=\left(\begin{array}{ll} 0&0\\ 0&k
\end{array}\right)\,\left(\begin{array}{c}
\omega\\ |\psi\rangle
\end{array}\right)=\left(\begin{array}{c} 0\\ |k\,\psi\rangle
\end{array}\right).
\end{equation}
The projection $P(s)$ has a form
\begin{equation} P(s)\, =\left(\begin{array}{ll} 1&0\\ 0&0
\end{array}\right),
\end{equation}
and the formal (reduced) resolvent $R(s)$ is given by
\begin{equation} R(s)=\left(\begin{array}{ll} 0&0\\ 0&k^{-1}
\end{array}\right).
\end{equation}
 The time dependence of this unitary family  can be encoded in the
rate of change of two operators, namely
\begin{equation} \dot P(s)=\left(\begin{array}{ll} 0&
\langle f_{s}|\\ |f_{s}\rangle&0
\end{array}\right)
\end{equation}
and
\begin{equation} (1-P(s))\dot{(\dot V(s) V^{\dagger}(s))} P(s)=\left(\begin{array}{ll}
 0&0\\ |g_{s}\rangle&0
\end{array}\right).
\end{equation}
 Suppose that $\dot V(s)
V^{\dagger}(s)$ is bounded, has a bounded derivative, and
\begin{equation}
\int_{B_I}| f_{s}|^{2}\, d\mu (k) \le O(|I|^{2\alpha}), \quad
\int_{B_I}|g_{s}|^{2}\, d\mu (k) \le O(|I|^{2\alpha}), \quad
\alpha\ge 0 ,\label{estim}
\end{equation}
where $B_I$ stands for a ball of radius $I$.
\begin{proposition}\label{friedrichs}
For the Friedrichs model described above, the evolution of the
state that starts as the bound state $\psi_{\tau}(0)\in Range
P(0)$,  is such that it remains close to the instantaneous bound
state and
\begin{equation}
dist\Big(\psi_\tau (s), Range P(s)\Big) \le
\left\{\begin{array}{lr}O\left(\frac{1}{\tau}\right),&\alpha >1;\\
 O\left(\frac{\log \tau}{\tau}\right),&\alpha =1;\\
O\left(\tau^{-\alpha}\right),&\alpha < 1,
\end{array}\right.
\end{equation}
 for all $s\in [0,1]$.
\end{proposition}

\begin{proof}
Formally
\begin{equation}
X= R\dot P +\dot P R=\left(\begin{array}{lr}
0&\left\langle\frac{f_s}{k}\right\vert\\
\left\vert\frac{f_s}{k}\right\rangle &0\end{array}\right)
\end{equation}
solves the commutator equation
\begin{equation}
[X,H]=[P,\dot P].\end{equation} Now choose
\begin{equation}
X_{\epsilon}= R_\epsilon\dot P +\dot P
R_\epsilon=\left(\begin{array}{lr}
0&\left\langle\frac{f_s\chi(k>\epsilon)}{k}\right\vert\\
\left\vert \frac{f_{s} \chi(k>\epsilon)}{k}\right\rangle
&0\end{array}\right),
\end{equation}
where $R_\epsilon=R(s)\chi(k>\epsilon)$ and pick $Y_{\epsilon}$
according to Eq.~(\ref{commutators}),
\begin{eqnarray}
Y_{\epsilon}&=&[P,\dot P]-[X_{\epsilon},H]\nonumber \\ &=&
\left(\begin{array}{lr} 0&\langle f_s\chi(k<\epsilon)\vert \\
\vert f_{s} \chi(k<\epsilon)\rangle&0\end{array}\right).
\end{eqnarray}
Then
\begin{equation}
\Vert Y_{\epsilon}\Vert \le O\left(\int_{0}^{\epsilon}|f_s|^2\,
d\mu (k)\right)\approx \epsilon^{2\alpha}.
\end{equation}
Since
\begin{equation}
\int_{\real^d/B_\epsilon}\frac{|f|^2}{k^{2}}\, d\mu(k) \le\left\{
\begin{array}{lr}
O(\epsilon^{2(\alpha-1)})&{\rm if}\  \alpha\not=1\\
-O(\log\epsilon)& {\rm if}\  \alpha =1
\end{array}\right.
\end{equation}
we get the appropriate estimate of $\Vert X_\epsilon\Vert$.
 What  remains is to estimate the norm of
$\dot{(X_{\epsilon}P(s))} P(s)$:
\begin{eqnarray}
\dot{(X_{\epsilon}P(s))} P(s) &=&
 R_{\epsilon}\ddot
P(s) P(s)+ \dot R_{\epsilon}\dot P(s) P(s)
 \nonumber \\ &=&
BR_{\epsilon}\dot P(s) P(s)-R_{\epsilon}\dot P(s)B
P(s)+R_{\epsilon}\dot B P(s)
 \nonumber \\ &=&
 BX_{\epsilon}P(s)-X_{\epsilon}B P(s)+R_{\epsilon}(1-P(s))\dot B P(s),
\end{eqnarray}
where $B=\dot V(s) V^{\dagger}(s)$. Making use of~(\ref{estim}) we
obtain that
\begin{equation}
\Vert X_{\epsilon}P(s)\Vert\, , \Vert\dot X_{\epsilon} P(s)\Vert\
\le \left\{
\begin{array}{lr}
O\left(\epsilon^{2(\alpha-1)}\right)&{\rm if}\  \alpha\not=1\\
-O\left(\log\epsilon\right)& {\rm if}\  \alpha =1
\end{array}\right. .
\end{equation}
So, provided $\alpha>1$, we get the adiabatic theorem with $Y=0$
and with a rate $1/\tau$. When $\alpha<1$ we
  optimize which gives
\begin{equation}
\frac{\Vert X_{\epsilon}P(s)\Vert+\Vert \dot X_{\epsilon}P(s)\Vert
}{\tau}+\Vert Y_{\epsilon}\Vert \le \left\{
\begin{array}{lr}
O\left(\frac{\epsilon^{\alpha-1}}{\tau}\right)+O\left(\epsilon^\alpha\right)&
{\rm if}\  \alpha<1\\
-O\left(\frac{\log\epsilon}{\tau}\right)+O(\epsilon^{\alpha})&
{\rm if}\ \alpha =1
\end{array}\right. .
\end{equation}
\qed
\end{proof}

\begin{acknowledgement} We are grateful to M. Aizenman for
suggesting using the regularity of measures to streamline the
proof of the main theorem, V. Bach,  R. Seiler and H. Spohn  for
useful discussions and hospitality.  This work was partially
supported by a grant from the Israel Academy of Sciences, the
Deutsche Forschungsgemeinschaft, and by the Fund for Promotion of
Research at the Technion.
\end{acknowledgement}

\makereferee{Communicated by B. Simon}

\begin{thebibliography}{article}
\bibitem{ah} Arai, A., Hirokawa, M.: {\it On the existence
 and uniqueness of ground
states of a generalized spin-boson model.} {J.\  Funct.\ Anal.\ }
{\textbf 151} (2), 455--503 (1997)
\bibitem{arnold} Arnold, V.I.: {\textit Geometrical Methods in
the theory of Ordinary Differential Equations.}
Berlin--Heidelberg--New-York: Springer, 1983
\bibitem{ae} Avron, J. E., Elgart, A.: {\it An adiabatic theorem
 without a gap condition:
 Two level system coupled to quantized radiation field.}
  {Phys.\ Rev.\ A\ } {\textbf 58,} 4300-4306 (1998)
\bibitem{ahs} Avron, J. E., Howland, J. S., Simon, B.: {\it Adiabatic
theorems for dense point spectra.} { Comm.\ in\ Math.\ Phys. \ }
{\textbf 128,} 497--507 (1990)
\bibitem{asy} Avron, J. E., Seiler, R., Yaffe, L. G.: {\it Adiabatic
theorems
 and applications to the quantum Hall effect.} { Comm.\ in\ Math.\ Phys. \ }
  {\textbf 110,} 33--49 (1987), (Erratum: { Comm.\ in\ Math.\ Phys. \ }
{\textbf 153,}  649-650 (1993))
\bibitem{bfs} Bach, V., Fr\"ohlich, J., Sigal, I. M.: {\it
Mathematical theory of nonrelativistic matter and radiation.}
{Lett.\ Math.\ Phys. \ } {\textbf 34,} 183--201 (1995)
\bibitem{bfs1} Bach, V., Fr\"ohlich, J., Sigal, I. M.: {\it Quantum
electrodynamics of confined non-relativistic particles.}  {Adv. \
in \ Math.} {\textbf 137,} 299 (1998)
\bibitem{bfs2} Bach, V., Fr\"ohlich, J., Sigal, I. M.: {\it Renormalization
group analysis of spectral problems in quantum field theory.}
{Adv. \ in \ Math.} {\textbf 137,} 205--298 (1998)
\bibitem{bfss} Bach, V., Fr\"ohlich J., Sigal, I. M., Sofer, A.:
 {it Positive commutators and spectrum of non-relativistic QED.} To appear
\bibitem{berry2} Berry, M.V.: {Proc.\ Roy.\ Soc.\ Lond. \ A \ } {\textbf 392,}
 45 (1984); The quantum phase: Five years after. In:  {\textit Geometric phases in
physics.} Shapere, A. and Wilczek, F., eds., Singapore: World
Scientific, 1989
\bibitem{berry} Berry, M.V.: {\it Histories of adiabatic transition.}
{Proc. Roy. Soc. Lond. A} {\textbf 429,} 61-72 (1990)
\bibitem{br} Berry, M.V., Robbins, J.M.: {\it Chaotic classical and
half classical adiabatic reactions: Geometric magnetism and
deterministic friction.} {Proc. Roy. Soc. Lond. A} {\textbf 442,}
659-672 (1993). {Proc.~Roy.~Soc.~A} {\textbf 392,} 45 (1984)
\bibitem{bs} Bethe, H.A., Salpeter, E.E.: {\textit Quantum Mechanics of
one and two electron atoms.} New York: Plenum, 1977
\bibitem{born} Born, M.: {\textit The Mechanics of the Atom.} New-York: Ungar, 1960
\bibitem{bf} Born, M., Fock, V.: {\it Beweis des Adiabatensatzes.} { Z. \
Phys. \ } {\textbf 51,} 165--169 1928
\bibitem{cohen} Cohen-Tannoudji, C.,
 Dupont-Roc, J.,  Grynberg G.:
{\textit Atoms and Photons Interactions.} New York: Wiley, 1992
\bibitem{combes} Combes, J. M.: In: {\textit Differential equations with
applications to mathematical physics.} Boston: Academic Press
1993; Combes, J. M., Montcho, R.: {\it Remarks on the relation
between quantum dynamics and fractal spectra.} {J. \ Math. \ Anal.
\ and \ Appl.\ } {\textbf 213,} 698--722 1997
\bibitem{cycon} Cycon, H. L., Froese, R.G., Kirsch, W., Simon, B.: {\textit
Schr\"odinger Operators.} Berlin--Heidelberg--New-York: Springer
1987
\bibitem{ds} Davies, E. B., Spohn, H.: {\it Open Quantum Systems with
Time-Dependent Hamiltonians and Their Linear Response.}
 { J.\ Stat.\ Phys.\ } {\textbf 19,} 511--523 1978
\bibitem{ehrenfest} Ehrenfest, P.: {\it Adabatische Invarianten u.
Quantentheorie.} {Ann. d. Phys.} {\textbf 51,} 327 1916
\bibitem{friedrichs1} Friedrichs, K.~O.: {\it On the perturbation of continuous
spectra.} {\it Comm.\ Pure \ Appl. \ Math. \ } {\textbf 1,}
361--406 1948
\bibitem{friedrichs2} Friedrichs, K.~O.: {\it Special topics in quantum theory.}
 Lecture notes, Courant Institute of Mathematical Science, New York University,
 (1953); {\it On the adiabatic theorem in quantum
theory}, Part I. Courant Institute of Mathematical Science, New
York University, (1955); {\it On the adiabatic theorem in quantum
theory}, Part II. Courant Institute of Mathematical Science, New
York University, 1956
\bibitem{g} Garrido, L. M.: {\it Generalized adiabatic invariance.}
 { J.\ Math.\ Phys.\ } {\textbf 5,} 355--362 1964
\bibitem{gp} Galindo, A., Pascual, P.: {\textit Quantum mechanics.}
Berlin--Heidelberg--New-York: Springer-Verlag, 1991
\bibitem{k} Golin, S., Knauf, A., Marmi, S.: {\it The Hannay angles:
Geometry, Adiabaticity and an example.} {Comm.\ in\ Math.\ Phys.\
} {\textbf 123,} 95-122 1989
\bibitem{guarneri} Guarneri, I.: {\it On the dynamical meaning of
spectral dimensions.}  { Ann.\ Inst.\ H.\ Poincar$\acute{e}$.} To
appear
\bibitem{hagedorn} Hagedorn, G.: {\it Adiabatic Expansions near
 Eigenvalue Crossings.} {Ann.\ Phys.} {\textbf 196,} 278-295 1989
\bibitem{hs} Huebner, Spohn, H.:  {Ann.\ Inst.\ H.
Poincar$\acute{e}$\ Phys.\ Theor. \ } {\textbf 62,} no. 3, 289
1995
\bibitem{js} Jak\u{s}i\'{c}, V., Segert, J.: {\it On the Landau Zener
formula for two-level systems.} {J. Math. Phys.} {\textbf 34,}
2807-2820 1993
\bibitem{jarzinski} Jarzinski, C.: {\it Multiple-time-scale approach
 to ergodic adiabatic systems: Another look.} {Phys. Rev. Lett.} {\textbf 71,}
  839 1993
\bibitem{joye} Joye, A., Pfister, C.E.: {\it Exponential Estimates in
Adiabatic Quantum Evolution.} Proceeding of the XII ICMP, Brisbane
Australia (1997); {\it Quantum Adiabatic Evolution.}  In {\textit
On Three Levels.} Fannes, M., Maes, C., Verbure, A., editors,
London: Plenum, 1994
\bibitem{kato1} Kato, T.: {\it Integration of the equation of evolution in a
 Banach space.} { J. \ Math. \ Soc.\ Japan.\ } {\textbf 5,} 208--234 1953
\bibitem{kato2} Kato, T.: {\it On the adiabatic theorem of quantum
mechanics.} { Phys.\ Soc.\ Jap.\ } {\textbf 5,} 435--439 1958
\bibitem{kato3} Kato, T.: {\textit Perturbation Theory for Linear
Operators.} Berlin--Heidelberg--New-York: Springer, 1966.
\bibitem{ks} Klein, M., Seiler, R.: {\it Power law corrections to the Kubo
formula vanish in quantum Hall systems.} { Comm.\ in \ Math.\
Phys. \ } {\textbf 128,} 141 1990
\bibitem{last} Last, Y.: {\it Quantum Dynamics and Decomposition of
Singular Continuous Spectra.} {J. Funct. Anal.} {\textbf 142,}
406-445 1996
\bibitem{lenard} Lennard, A.: {\it Adiabatic Invariance to All
Orders.} {Ann. Phys.} {\textbf 6,} 261-276 1959
\bibitem{lm} Lochak, P., Meunier, C.: {\textit Multiphase Averaging for
Classical systems.} Berlin--Heidelberg--New-York: Springer,  1988
\bibitem{martinez} Martinez, A.: {\it Precise exponential estimates in
adiabatic theory}, {J. Math. Phys.} {\textbf 35,} 3889-3915 1994
\bibitem{mn} Martinez, A., Nakamura, S.: {\it Adiabatic limit and
scattering.} {C.R. Acd. Sci. Paris.} {\textbf 318,} 1153-1158 1994
\bibitem{nt} Narnhofer, H., Thirring, W.: {\it Adiabatic theorem in
quantum statistical mechanics.}  {Phys. Rev. A \ } {\textbf 26,}
3646, 1982
\bibitem{n} Nenciu, G.: {\it On the adiabatic theorem of quantum mechanics.}
{J.\ Phys.\ A} {\textbf 13,} L15-L18 1980
\bibitem{N} Nenciu, G.: {\it Linear Adiabatic Theory: Exponential
Estimates.} {Comm.\ in\ Math.\ Phys.\ } {\textbf 152,} 479-496
1993
\bibitem{ott} Ott, E.: {\it Goodness of ergodic adiabatic invariants.} { Phys.\
Rev.\ Lett.\ } {\textbf 42,} 1628-1631 1979; and
 Brown R., Ott, E., Grebogi, C.: {\it Goodness of ergodic adiabatic
invariants.} { J.\ Stat.\ Phys.\ } {\textbf 49,} 511-550 1987
\bibitem{rs} Reed, M., Simon, B.: {\textit Methods of Modern
Mathematical Physics II. Fourier Analysis, Self-Adjointness.}
London: Academic Press, 1975
 \bibitem{riemann} Riemann, B.: {\it Ueber der Darstellbarkeit einer
 Function durch  einen tri\-go\-no\-met\-rishe Reihe.}
 In: { Math.\ Werke}, Leipzig: Teubner, pp. 213--253 1876;
 Lebesgue, H.: {\it Sur les S$\acute{e}$ries Trigonom$\acute{e}$triques.}
 {Ann.\ Sci.\ Ecole\ Norm.\ Sup.\ } {\textbf 20,} 453--485 1903
\bibitem{simon} Simon, B.: {\it The
theory of resonances for dilation analytic potentials and
 the foundations of time-dependent
perturbation theory.} {Ann. of Math.} {\textbf 97,} 247-274 1973
\bibitem{s} Spohn, H.: {\it Ground state(s) of the spin-boson Hamiltonian.}
 {Comm.\ Math.\ Phys.\ } {\textbf 123,} 277--304 1989
\bibitem{thouless} Thouless, D.J.: {\textit Topological Quantum Numbers in
Nonrelativistic Phy\-sics.} Singapore: World Scientific, 1998
\bibitem{y} Yosida, K.: {\textit Functional Analysis.} Berlin:
 Springer-Verlag, 1968
\end{thebibliography}
\end{document}